\pdfoutput=1
\def\year{2018}\relax
\documentclass[letterpaper]{article} 
\usepackage{aaai18}  
\usepackage{times}  
\usepackage{helvet}  
\usepackage{courier}  
\usepackage{url}  
\usepackage{graphicx}  

\usepackage{booktabs} 
\usepackage{amsmath}
\usepackage{amsfonts}
\usepackage{multirow}
\usepackage{url}
\usepackage{caption}
\usepackage{subcaption}

\DeclareMathOperator*{\argmax}{arg\,max}

\newcommand{\newcite}[1]{\citeauthor{#1}~\shortcite{#1}}

\begin{document}
%
 \title{Contrastive Training for Models of Information Cascades}

 \author{
  Shaobin Xu \and David A. Smith \\
 College of Computer and Information Science\\Northeastern University\\
 440 Huntington Avenue\\
 Boston, MA, 02115\\
 \texttt{\{shaobinx,dasmith\}@ccs.neu.edu}
 }

\maketitle
\begin{abstract}
  This paper proposes a model of information cascades as directed
  spanning trees (DSTs) over observed documents.  In addition, we
  propose a contrastive training procedure that exploits partial
  temporal ordering of node infections in lieu of labeled training
  links.  This combination of model and unsupervised training makes it
  possible to improve on models that use infection times alone and to
  exploit arbitrary features of the nodes and of the text content of
  messages in information cascades.  With only basic node and time lag
  features similar to previous models, the DST model achieves
  performance with unsupervised training comparable to strong
  baselines on a blog network inference task.  Unsupervised training
  with additional content features achieves significantly better
  results, reaching half the accuracy of a fully supervised model.
\end{abstract}

\section{Introduction}

As is their wont, politicians talk---on television, on the floor of
the legislature, in printed quotations, and on their websites and
social media feeds.  If you read and listen to all these statements,
you might notice common tropes and turns of phrase that groups of
politicians used to describe some issue \cite{grimmer13:politanal}.
You might even discover a list of ``talking points'' underlying this
common behavior.\footnote{The U.S. State Department, for example,
  produced a much-discussed set of talking points memos in response to
  the 2012 attack in Benghazi.}  Similarly, you might be reading the
literature in a scientific field and find that a paper from another
discipline starts getting cited repeatedly.  Which previous paper, or
papers, introduced the new technique?  Or perhaps you read several
news stories about a new product from some company and then find that
they all share text with a press release put out by the company.

In each of these cases, we might be interested in structures at
differing levels of detail.  We might be interested in individual
\textbf{links}, e.g., knowing which previous paper it was that later
papers were mining for further citations; in \textbf{cascades}, e.g.,
knowing which news stories are copying from which press releases or
from each other; and in \textbf{networks}, e.g., knowing which
politicians are most likely to share talking points or which
newspapers are most likely to publish press releases from particular
businesses or universities.  Depending on our data source, some of
these structures could be directly observed.  With the right API
calls, we might observe retweets (links), chains of retweets
(cascades), and follower relations (networks) on Twitter.  We might
also be interested in inferring an underlying social network for which
the Twitter follower relation is partial evidence.  In contrast,
politicians interviewed on television do not explicitly cite the
sources of their talking points, which must be inferred.

Observing the diffusion process often reduces to keeping track of when
nodes (newspapers, bills, people, etc.) mention a piece of
information, reuse a text, get infected, or exhibit a
\textit{contagion} in a general sense.  When the structure of the
propagation of contagion is hidden and we cannot tell which node
infected which, all we have is the result of diffusion process---that
is, the timestamp and possibly other information when the nodes get
infected.  We want to infer the diffusion process itself by using such
information to predict the links of underlying network. There have
been increasing efforts to uncover and model different types of
information cascades on networks \cite{2016arXiv161000782B}: modeling
hidden networks from observed infections
\cite{stack2012inferring,rodriguez2014uncovering}, modeling topic
diffusion in networks \cite{gui2014modeling}, predicting social
influence on individual mobility \cite{mastrandrea2015contact} and so
on.

This work all focuses on using parametric models of the time
differences between infections.  Such models are useful when the only
information we can get from the result of diffusion process is the
timestamps of infections.  We can hope to make better predictions,
however, with access to additional features, such as the location of
each node, the similarity between the messages received by two nodes,
etc.  Popular parametric models cannot incorporate these features into
unsupervised training.

In this paper, we propose an edge-factored, conditional log-linear
directed spanning tree (DST) model with an unsupervised, contrastive
training procedure to infer the link structure of information
cascades.  After reviewing related work, we describe the DST model in
detail, including an efficient inference algorithm using the directed
matrix-tree theorem and the gradient of our maximum conditional
likelihood optimization problem.  We then report experiments on the
ICWSM Spinn3r dataset, where we can observe the true hyperlink
structure for evaluation and compare the proposed method to
MultiTree \cite{rodriguez2012submodular} and InfoPath
\cite{rodriguez2014uncovering} and some simple, but effective, baselines.
We conclude by discussing directions for future work.

\section{Related Work}
\label{related-work}

There has been a great deal of work on trying to infer underlying
network structure using information cascades, most of which are based
on the independent cascade (IC) model \cite{saito2008prediction}.  We
evaluate the DST model against a transmission based model from
\newcite{rodriguez2012submodular}, which, similar to our work, also
uses directed spanning trees to represent cascades, but employs a
submodular parameter-optimization method and fixed activation rates.
In addition, we compare our work with an advanced model from
\newcite{rodriguez2014uncovering}, which uses a generative
probabilistic model for inferring both static and dynamic diffusion
networks. It is a line of work starting from using a generative model
with fixed activation rate
\cite{gomez2010inferring,rodriguez2012submodular,myers2010convexity}.
Later comes the development of inferring the activation rate between
nodes to reveal the network structure
\cite{rodriguez2011uncovering,gomez2013structure,snowsill2011refining,rodriguez2013modeling,rodriguez2014uncovering}.
\newcite{zhai2015cascade} use a Markov chain Monte Carlo approach for
the inference problem.  \newcite{linderman2014discovering} propose a
probabilistic model based on mutually-interacting point processes and
also use MCMC for the inference.  \newcite{gui2014modeling} model
topic diffusion in multi-relational networks.  An interesting approach
by \newcite{amin2014learning} infers the unknown network structure,
assuming the detailed timestamps for the spread of the contagion are
not observed but that ``seeds'' for cascades can be identified or even
induced experimentally.  \newcite{wang2012feature} propose
feature-enhanced probabilistic models for diffusion network inference
while still maintaining the requirement that exact propagation times
be observed and modeled.  \newcite{daneshmand2014estimating} and
\newcite{abrahao2013trace} perform theoretical analysis of
transmission-based cascade inference models.  While the foregoing
approaches are all based on parametric models of propagation time
between infections, \newcite{rong2016model} experiment with a
nonparametric approach to discriminating the distribution of diffusion
times between connected and unconnected nodes.  Recently,
\newcite{2016arXiv161000782B} have compiled a survey about the methods
and applications for different network structure inference problems.

Tutte's directed matrix-tree theorem, which plays a key role in our
approach, has been used in natural language processing to infer
posterior probabilities for edges in nonprojective syntactic
dependency trees
\cite{Smith:2007ts,koo2007structured,mcdonald2007complexity} and for
inferring semantic hierarchies (i.e., ontologies) over words
\cite{bansal14:acl}.

\section{Method}
\label{method}

In this section, we present our modeling and inference approaches.  We
first present a simple log-linear, edge-factored directed spanning
tree (DST) model of cascades over network nodes.  This allows us to
talk concretely about the likelihood objective for supervised and
unsupervised training, where we present a \textbf{contrastive}
objective function.  We note that other models besides the DST model
could be trained with this contrastive objective.  Finally, we derive
the gradient of this objective and its efficient computation using
Tutte's directed matrix-tree theorem.

\subsection{Log-linear Directed Spanning Tree Model}
\label{method:log-linear-model}

For each cascade, define a set of activated nodes
$\mathbf{x}=\{ x_1,\dots, x_n \}$, each of which might be associated
with a timestamp and other information that are the input to the
model.  Nodes thus correspond to (potentially) dateable entities such
as webpages or posts, and not aggregates, such as websites or users.
Let $\mathbf{y}$ be a directed spanning tree of $\mathbf{x}$, which is
a map $\mathbf{y}:\{1,\dots,n\} \rightarrow \{0, 1, \dots, n\}$ from
child indices to parent indices of the cascade.  In the range of
mapping $\mathbf{y}$ we add a new index $0$, which represents a dummy
``root'' node $x_0$.  This allows us to model both single cascades and
to disentangle multiple cascades on a set of nodes $\mathbf{x}$,
since more than one ``seed'' node might attach to the dummy root.  In
the experiments below, we model datasets with both single-rooted
(``separated'') and multi-rooted (``merged'') cascades.

A valid directed spanning tree is by definition acyclic.  Every node
has exactly one parent, with the edge
$x_{\mathbf{y}(i)}\rightarrow x_i$, except that the root node has
in-degree 0.  We might wish to impose additional constraints $C$ on
the set of spanning trees: for instance, we might require that edges
not connect nodes with timestamps known to be in reverse order.  Let
$\mathcal{Y}_C$ be the set of all valid directed spanning trees that
satisfy a list of rules in constraint set $C$ over $\mathbf{x}$, and
$\mathcal{Y}$ be the set of all directed spanning trees over the same
sequence of $\mathbf{x}$ but without any constraint being imposed.

Define a log-linear model for trees over $\mathbf{x}$.  The
unnormalized probability of the tree $\mathbf{y} \in \mathcal{Y}$ is
thus:

\begin{equation}
  \label{eq:log-linear}
  \tilde{p}_{\vec{\theta}}(\mathbf{y} \mid \mathbf{x}) = e^{\vec{\theta} \cdot \vec{f}(\mathbf{x},\mathbf{y})}
\end{equation}

\noindent where $\vec{f}$ is a feature vector function on cascade and
$\vec{\theta} \in \mathbb{R}^m$ parameterizes the model.  Following
\cite{mcdonald05:acl}, we assume that features are
\textbf{edge-factored}:

\begin{equation}
  \label{eq:edge-factored}
  \vec{\theta} \cdot \vec{f}(\mathbf{x},\mathbf{y})=
  \sum_{i=1}^{n} \vec{\theta} \cdot \vec{f}_{\mathbf{x}}(\mathbf{y}(i), i) = \sum_{i=1}^n s(\mathbf{y}(i), i)
\end{equation}

\noindent where $s(i, j)$ is the \emph{score} of a directed edge
$i \rightarrow j$.  In other words, given the sequence $\mathbf{x}$
and the cascade is a directed spanning tree, this directed spanning
tree (DST) model assumes that the edges in the tree are all
conditionally independent of each other.

Despite the constraints they impose on features, we can perform
inference with edge-factored models using tractable $O(n^3)$
algorithms, which is one of the advantages this model brings.  Since
$\tilde{p}_{\vec{\theta}}(\mathbf{y} \mid \mathbf{x})$ is not a
normalized probability, we divide it by the sum over all possible
directed spanning trees, which gives us:

\begin{equation}\label{eq:like}
  p_{\vec{\theta}}(\mathbf{y} \mid \mathbf{x}) = \frac{e^{\vec{\theta} \cdot \vec{f}(\mathbf{x},\mathbf{y})}}
  {\sum_{\mathbf{y'} \in \mathcal{Y}} e^{\vec{\theta} \cdot \vec{f}(\mathbf{x},\mathbf{y'})}}
  = \frac{e^{\vec{\theta} \cdot \vec{f}(\mathbf{x},\mathbf{y})}}{Z_{\vec{\theta}}(\mathbf{x})}
\end{equation}

\noindent where $Z_{\vec{\theta}}(\mathbf{x})$ denotes the sum of
log-linear scores of all directed spanning trees, i.e., the partition
function.

If, for a given set of parameters $\vec{\theta}$, we merely wish to
find the
$\mathbf{\hat y} = \argmax_{\mathbf{y} \in \mathcal{Y}}
p_{\vec{\theta}}(\mathbf{y} \mid \mathbf{x})$, we can pass the scores
for each $i \rightarrow j$ candidate edge to the Chu-Liu-Edmonds
maximum directed spanning tree algorithm \cite{chu65,edmonds67}.

\subsection{Likelihood of a cascade}
\label{sec:likelihood}

When we observe all the directed links in a training set of cascades,
we now have the machinery to perform supervised training with maximum
conditional likelihood.  We can simply maximize the likelihood of
the true directed spanning tree
$p_{\vec\theta}(\mathbf{y^*} \mid \mathbf{x})$ for each cascade in our
training set, using the gradient computations discussed below.

When we do not observe the true links in a cascade, we need a
different objective function.  While we cannot restrict the numerator
in the likelihood function (\ref{eq:like}) to a single, true tree, we
can restrict it to the set of trees $\mathcal{Y}_C$ that obey some
constraints $C$ on valid cascades.  As mentioned above, these
constraints might, for instance, require that links point forward in
time or avoid long gaps.  We can now write the likelihood function for
each cascade $\mathbf{x}$ as a sum of the probabilities of all
directed spanning trees that meet the constraints $C$:

\begin{equation}
  \label{method:likelihood-func}
  \mathcal{L}_\mathbf{x}
  = \sum_{\mathbf{y} \in \mathcal{Y}_\mathbf{C}} p_{\vec{\theta}}(\mathbf{y} \mid \mathbf{x}) 
  = \frac{\sum_{\mathbf{y} \in \mathcal{Y}_\mathbf{C}} e^{\vec{\theta} \cdot \vec{f}(\mathbf{x},\mathbf{y})}}
  {Z_{\vec{\theta}}(\mathbf{x})} 
  = \frac{Z_{\vec{\theta},C}(\mathbf{x})}{Z_{\vec{\theta}}(\mathbf{x})}
\end{equation}

\noindent where $Z_{\vec{\theta},C}(\mathbf{x})$ denotes the sum of
log-linear scores of all valid directed spanning trees under
constraint set $C$.

This is a \textbf{contrastive} objective function that, instead of
maximizing the likelihood of a single outcome, maximizes the
likelihood of a \textbf{neighborhood} of possible outcomes contrasted
with implicit negative evidence \cite{Smith05ContrastiveACL43}.  A
similar objective could be used to train other cascade models besides
the log-linear the DST model presented above, e.g., models such as the
Hawkes process in \newcite{linderman2014discovering}.

As noted above, cascades on a given set of nodes are assumed to be
independent.  We thus have a log-likelihood over all $N$ cascades:

\begin{equation}
  \log \mathcal{L}_\mathbf{N} = \sum_{\mathbf{x}} \log \mathcal{L}_{\mathbf{x}}
  = \sum_{\mathbf{x}} \log \frac{Z_{\vec{\theta},C}(\mathbf{x})}{Z_{\vec{\theta}}(\mathbf{x})}
\end{equation}

\subsection{Maximizing Likelihood}
\label{sec:max-like}

Our goal is to find the
parameters $\vec{\theta}$ that solve the following maximization problem:
problem:
\begin{equation}
  \max_{\vec{\theta}} \log \mathcal{L}_N = \max_{\vec{\theta}} \sum_{\mathbf{x}}
  (\log Z_{\vec{\theta},C}(\mathbf{x}) - \log Z_{\vec{\theta}}(\mathbf{x}))
\end{equation}

\noindent To solve this problem with quasi-Newton numerical
optimization methods such as L-BFGS \cite{liu1989limited}, we need to
compute the gradient of the objective function, which for a given
parameter $\theta_k$ is given by the following equation:

\begin{equation}
  \frac{\partial \log \mathcal{L}_N}{\partial \theta_k} = \sum_\mathbf{x} \left(
    \frac{\partial \log Z_{\vec{\theta},C}(\mathbf{x})}{\partial \theta_k}
  - \frac{\partial \log Z_{\vec{\theta}}(\mathbf{x})}{\partial \theta_k} \right)
\end{equation}

For a cascade that contains $n$ nodes, even if we have tractable
number of valid directed spanning trees in $\mathcal{Y}_C$, there will
be $n^{n-2}$ (Cayley's formula) possible directed spanning trees for
the normalization factor $Z_{\vec{\theta}}(\mathbf{x})$, which makes
the computation intractable.  Fortunately, there exists an efficient
algorithm that can compute $Z_{\vec{\theta}}(\mathbf{x})$, or
$Z_{\vec{\theta},C}(\mathbf{x})$, in $O(n^3)$ time.

\begin{table*}[t]
\centering
\caption{Cascade-level inference of DST with different feature sets, in unsupervised
  learning setting (Table~\ref{tbl:cascade-infer-unsup}) in comparison with naive
  attach-everything-to-earliest baseline, as well as supervised learning setting
  (Table~\ref{tbl:sup-merged})} 
\label{tbl:cascade-infer}
\def\arraystretch{1.2}
\begin{subtable}[]{\linewidth}
\centering
\begin{tabular}{|c|l|l|c|c|c|}
\hline
Dataset & Cascade types & Method & Recall & Precision & F1 \\ \hline
  \multirow{6}{*}{ICWSM 2011} & \multirow{3}{*}{Separated Cascades} & DST Basic & 0.348 & 0.454 & 0.394 \\ \cline{3-6}
  & & DST Enhanced & \textbf{0.504} & \textbf{0.658} & \textbf{0.571} \\ \cline{3-6}
  & & Naive Baseline & 0.450 & 0.587 & 0.509 \\ \cline{2-6}
  & \multirow{3}{*}{Merged Cascades} & DST Basic & 0.027 & 0.035 & 0.031 \\ \cline{3-6}
  & & DST Enhanced & \textbf{0.036} & \textbf{0.047} & \textbf{0.040} \\ \cline{3-6}
 & & Naive Baseline & 0.015 & 0.019 & 0.017 \\ \hline
  \multirow{6}{*}{\begin{tabular}[|c|]{@{}c@{}}ICWSM 2011\\ (tree structure enforced)\end{tabular}} & \multicolumn{1}{|c|}{\multirow{3}{*}{Separated Cascades}} & DST Basic & 0.622 & 0.622 & 0.622 \\ \cline{3-6}
  & \multicolumn{1}{|c|}{} & DST Enhanced & \textbf{0.946} & \textbf{0.946} & \textbf{0.946} \\ \cline{3-6}
  & \multicolumn{1}{|c|}{} & Naive Baseline & 0.941 & 0.933 & 0.937 \\ \cline{2-6}
  & \multirow{3}{*}{Merged Cascades} & DST Basic & 0.042 & 0.042 & 0.042 \\ \cline{3-6}
  & & DST Enhanced & \textbf{0.246} & \textbf{0.246} & \textbf{0.246} \\ \cline{3-6}
 & & Naive Baseline & 0.043 & 0.043 & 0.043 \\ \hline
\end{tabular}
\caption{Unsupervised Setting}\label{tbl:cascade-infer-unsup}
\end{subtable}
\begin{subtable}[]{\textwidth}
\centering
\begin{tabular}{|l|c|c|c|c|}
  \hline
  \multirow{2}{*}{Merged Cascades} & \multicolumn{2}{c|}{Training} & \multicolumn{2}{c|}{Test} \\ \cline{2-5} 
   & Recall & Precision & Recall & Precision \\ \hline
  Basic Feature Set & 0.171$\pm$0.001 & 0.171$\pm$0.001 & 0.164$\pm$0.007 & 0.164$\pm$0.007 \\ \hline
  Enhanced Feature Set & 0.475$\pm$0.002 & 0.475$\pm$0.002 & 0.455$\pm$0.011 & 0.455$\pm$0.011 \\ \hline
  Naive Baseline & 0.042$\pm$0.001 & 0.042$\pm$0.001 & 0.046$\pm$0.009 & 0.046$\pm$0.009 \\ \hline
\end{tabular}
\caption{Supervised Setting}\label{tbl:sup-merged}
\end{subtable}
\end{table*}

\subsection{Matrix-Tree Theorem and Laplacian Matrix}
\label{sec:mtt}

\newcite{tutte1984graph} proves that for a set of nodes $x_0, \dots, x_n$, the
sum of scores of all directed spanning trees $Z_{\vec{\theta}}(\mathbf{x})$ 
in $\mathcal{Y}$ rooted at $x_j$ is
\begin{equation}
  \label{eq:matrix-tree-theorem}
  Z_{\vec{\theta}}(\mathbf{x}) = \left| \hat{\mathbf{L}}_{\vec{\theta},\mathbf{x}}^j \right|
\end{equation}
where $\hat{\mathbf{L}}_{\vec{\theta},\mathbf{x}}^j$ is the matrix produced by deleting
the $j$-th row and column from Laplacian matrix $\mathbf{L}_{\vec{\theta}, \mathbf{x}}$.


Before we define Laplacian matrix, we first denote:
\begin{equation}
  u_{\vec{\theta}, \mathbf{x}}(j, i) =
  e^{\vec{\theta} \cdot \vec{f}_\mathbf{x}(j, i)} = e^{s_{\vec\theta,\mathbf{x}}(j, i)}
\end{equation}
\noindent where $j = \mathbf{y}(i)$, which is the parent of $x_i$ in $\mathbf{y}$.
Recall that we define the unnormalized score of a spanning tree over $\mathbf{x}$
as a log-linear model using edge-factored scores (Eq~\ref{eq:log-linear},
\ref{eq:edge-factored}). Therefore, we have:
\begin{equation}
  e^{\vec{\theta}\cdot\vec{f}(\mathbf{x},\mathbf{y})}
  = e^{\sum_i \vec{\theta}\cdot\vec{f}_\mathbf{x}(\mathbf{y}(i), i)}
  = \prod_{i=1}^{n} u_{\vec{\theta},\mathbf{x}}(j, i)
\end{equation}

\noindent where $u_{\vec{\theta},\mathbf{x}}(j,i)$ represents the multiplicative contribution of the
edge from parent $j$ to child $i$ to the total score of the tree.

Now we can define the Laplacian matrix
$\mathbf{L}_{\vec{\theta},\mathbf{x}} \in \mathbb{R}^{(n+1) \times (n+1)}$
for directed spanning trees by:
\begin{equation}
  \begin{aligned}
    &\lbrack\mathbf{L}_{\vec{\theta},\mathbf{x}}\rbrack_{j,i}=\\
    &\left\{
      \begin{array}{rl}
        -u_{\vec{\theta},\mathbf{x}}(j,i) & \text{if edge } (j, i) \in C\\
        \sum\limits_{k \in \{0,\dots,n\}, k \ne j} u_{\vec{\theta},\mathbf{x}}(k,i) & \text{if  } j = i\\
                                                    0 & \text{if edge } (j,i) \notin C
      \end{array}
    \right.
  \end{aligned}
\end{equation}

\noindent where $j$ represents a parent node and $i$ represents a child node.
As for all possible valid directed spanning trees, we will have 0 for all entries
where the edge from parent $j$ to child $i$ doesn't satisfy the specified
constraint set. For all possible directed spanning trees, however, the
constraint set $C$ is $V \times V$, that is all possible edges.

We can use the LU factorization to compute the matrix inverse, so that
the determinant of the Laplacian matrix can be done in $O(n^3)$
times. Meanwhile, the Laplacian matrix is diagonally dominant, in that
we use positive edge scores to create the matrix.  The matrix
therefore is guaranteed to be invertible.

\subsection{Gradient}

\newcite{Smith:2007ts} use a similar inference approach for probabilistic models of
nonprojective dependency trees.  They derive that for any parameter
$\theta_k$,

\begin{equation}
  \label{eq:gradient-1}
  \begin{aligned}
    \frac{\partial \log Z_{\vec{\theta}}(\mathbf{x})}{\partial \theta_k}
    = \frac{1}{\left| \mathbf{L}_{\vec{\theta},\mathbf{x}} \right|}
    \sum_{i=1}^{n} \sum_{j=0}^{n} u_{\vec{\theta},\mathbf{x}}(j,i)f_\mathbf{x}^k(j,i)\\
    \times (\frac{\partial \left| \mathbf{L}_{\vec{\theta},\mathbf{x}} \right|}
    {\partial \lbrack \mathbf{L}_{\vec{\theta},\mathbf{x}} \rbrack_{i,i}}
    - \frac{\partial \left| \mathbf{L}_{\vec{\theta},\mathbf{x}} \right|}
    {\partial \lbrack \mathbf{L}_{\vec{\theta},\mathbf{x}} \rbrack_{j,i}})
  \end{aligned}
\end{equation}

\noindent Also, for an arbitrary matrix $\mathbf{A}$, they derive the
gradient of $\mathbf{A}$ with respect to any cell
$\lbrack \mathbf{A} \rbrack_{j,i}$ using the
determinant and entries in the inverse matrix:
\begin{equation}
  \label{eq:gradient-2}
  \frac{\partial \left| \mathbf{A} \right|}{\partial \lbrack \mathbf{A} \rbrack_{j,i}} =
  \left| \mathbf{A} \right| \lbrack \mathbf{A}^{-1} \rbrack_{i,j}
\end{equation}

Plugging (\ref{eq:gradient-2}) to (\ref{eq:gradient-1}) gives us the final gradient
of $Z_{\vec{\theta}}(\mathbf{x})$ with respect to $\theta_k$:
\begin{equation}
  \begin{aligned}
    \frac{\partial \log Z_{\vec{\theta}}(\mathbf{x})}{\partial \theta_k}
    = \sum_{i=1}^{n} \sum_{j=0}^{n} u_{\vec{\theta},\mathbf{x}}(j,i)f_\mathbf{x}^k(j,i)\\
    \times
    \left( \lbrack \mathbf{L}_{\vec{\theta},\mathbf{x}}^{-1} \rbrack_{i,i}
    - \lbrack \mathbf{L}_{\vec{\theta},\mathbf{x}}^{-1} \rbrack_{i,j} \right)
  \end{aligned}
\end{equation}

\begin{table*}[t]
\centering
\caption{Comparison of MultiTree, InfoPath and DST on inferring a static network
  on the original ICWSM 2011 dataset and on a dataset with enforced
  tree structure.  The DST model is trained and tested unsupervisedly
  on both separate cascades and merged cascades using different
  feature sets and the naive attach-everything-to-earliest-node
  baseline.}
\label{tbl:network-infer}
  \def\arraystretch{1.2}
\begin{tabular}{|c|l|l|c|c|c|c|}
\hline
  Dataset & Cascade types & Method & Recall & Precision & F1 & AP \\ \hline
  \multirow{8}{*}{ICWSM 2011} & \multirow{5}{*}{Separated Cascades} & MultiTree & 0.367 & 0.242 & 0.292 & N/A \\ \cline{3-7}
  & & InfoPath & 0.414 & 0.273 & 0.329 & N/A \\ \cline{3-7}
  & & DST Basic & 0.557 & 0.368 & 0.443 & 0.279 \\ \cline{3-7}
  & & DST Enhanced & \textbf{0.842} & 0.556 & \textbf{0.670} & \textbf{0.599}\\ \cline{3-7}
  & & Naive Baseline & 0.622 & \textbf{0.595} & 0.608 & 0.385\\ \cline{2-7}
  & \multirow{3}{*}{Merged Cascades} & DST Basic & 0.052 & 0.034 & 0.041 & \textbf{0.003} \\ \cline{3-7}
  & & DST Enhanced & \textbf{0.057} & \textbf{0.038} & \textbf{0.045} & \textbf{0.003} \\ \cline{3-7}
  & & Naive Baseline & 0.015 & 0.019 & 0.017 & 0.001\\ \hline
  \multirow{8}{*}{\begin{tabular}[|c|]{@{}c@{}}ICWSM 2011\\ (tree structure enforced)\end{tabular}} & \multirow{5}{*}{Separated Cascades} & MultiTree & 0.249 & 0.196 & 0.220 & N/A \\ \cline{3-7}
  & & InfoPath & 0.375 & 0.294 & 0.330 & N/A \\ \cline{3-7}
  & & DST Basic & 0.618 & 0.486 & 0.544 & 0.452 \\ \cline{3-7}
  & & DST Enhanced & \textbf{0.950} & 0.747 & \textbf{0.836} & \textbf{0.915} \\ \cline{3-7}
  & & Naive Baseline & 0.941 & \textbf{0.933} & 0.937 & 0.892\\ \cline{2-7}
  & \multirow{3}{*}{Merged Cascades} & DST Basic & 0.083 & 0.065 & 0.073 & 0.012\\ \cline{3-7}
  & & DST Enhanced & \textbf{0.207} & \textbf{0.163} & \textbf{0.182} & \textbf{0.047} \\ \cline{3-7}
  & & Naive Baseline & 0.043 & 0.043 & 0.043 & 0.005\\ \hline
\end{tabular}
\end{table*}

\section{Experiments}
\label{exp:blog}

One of the hardest tasks in network inference problems is gathering
information about the true network structure. Most existing work has
conducted experiments on both synthetic data with different parameter
settings and on real-world networks that match the assumptions of
proposed method.  Generating synthetic data, however, is less feasible
if we want to exploit complex textual features, which negates one of
the advantages of the DST model.  Generating child text from parent
documents is beyond the scope of this paper, although we believe it to
be a promising direction for future work.  In this paper, therefore,
we train and test on documents from the ICWSM 2011 Spinn3r dataset
\cite{burton2011icwsm}.  This allows us to compare our method with
MultiTree \cite{rodriguez2012submodular} and InfoPath
\cite{rodriguez2014uncovering}, both of which output a network given
a set of cascades. We also analyze the performance of DST at the
cascade level, an ability that MultiTree, InfoPath and similar methods lack.

\subsection{Dataset Description}
\label{icwsm:desc}

The ICWSM 2011 Spinn3r dataset consists of 386 million different web
posts, such as blog posts, news articles, social media content, etc.,
made between January 13 and February 14, 2011. We first avoid
including hyperlinks that connect two posts from the same websites, as
they could simply be intra-website navigation links.  In addition, we
enforce a strict chronological order from source post to destination
post to filter out erroneous date fields. Then, by backtracing
hyperlinks to the earliest ancestors and computing connected
components, we are able to obtain about 75 million clusters, each of
which serves as a separate cascade.  We only keep the cascades
containing between 5 and 100 posts, inclusive.  This yields 22,904
cascades containing 205,234 posts from 61,364 distinct websites.  We
create ground truth for each cascade by using the hyperlinks as a
proxy for real information flow. For time-varying network, we include
edges only appear in a particular day into the corresponding network,
while for the static network we simply include any existing edge
regardless of the timestamp.

Of these cascades, approximately 61\% don't have a tree structure, and
among the remaining cascades, 84\% have flat structures, meaning for
each cascade, all nodes other than the earliest one are in fact
attached to the earliest node in the ground truth. In this paper we
report the performance of our models on the original dataset and on a
dataset where the cascades are guaranteed to be trees.  To construct
the tree cascades, before finding the connected components using
hyperlinks, we remove posts which have hyperlinks coming from more
than one source website.  Selecting, as above, cascades whose sizes
are between 5 and 100 nodes, this yields 20,424 separate cascades
containing 201,875 posts from 63,576 distinct websites

We also merge cascades to combine all cascades that start within an
hour of each other to make the inference problem more challenging and
realistic, since if we don't know links, we are unlikely to know the
membership of nodes in cascades exactly. In the original ICWSM 2011
dataset, we obtain 789 merged cascades and for the tree-constrained
data, we obtain 938 merged cascades.  When merging cascades, we only
change the links to the dummy root node, and the underlying network
structure remains the same. The DST model would be able to learn
different parameters depending on whether we train it on separate
cascades or merged cascades. We report on the comparison between both
with MultiTree and InfoPath.

\begin{figure*}[t]
	\centering
  \begin{subfigure}[t]{0.33\textwidth}
    \includegraphics[width=\textwidth]{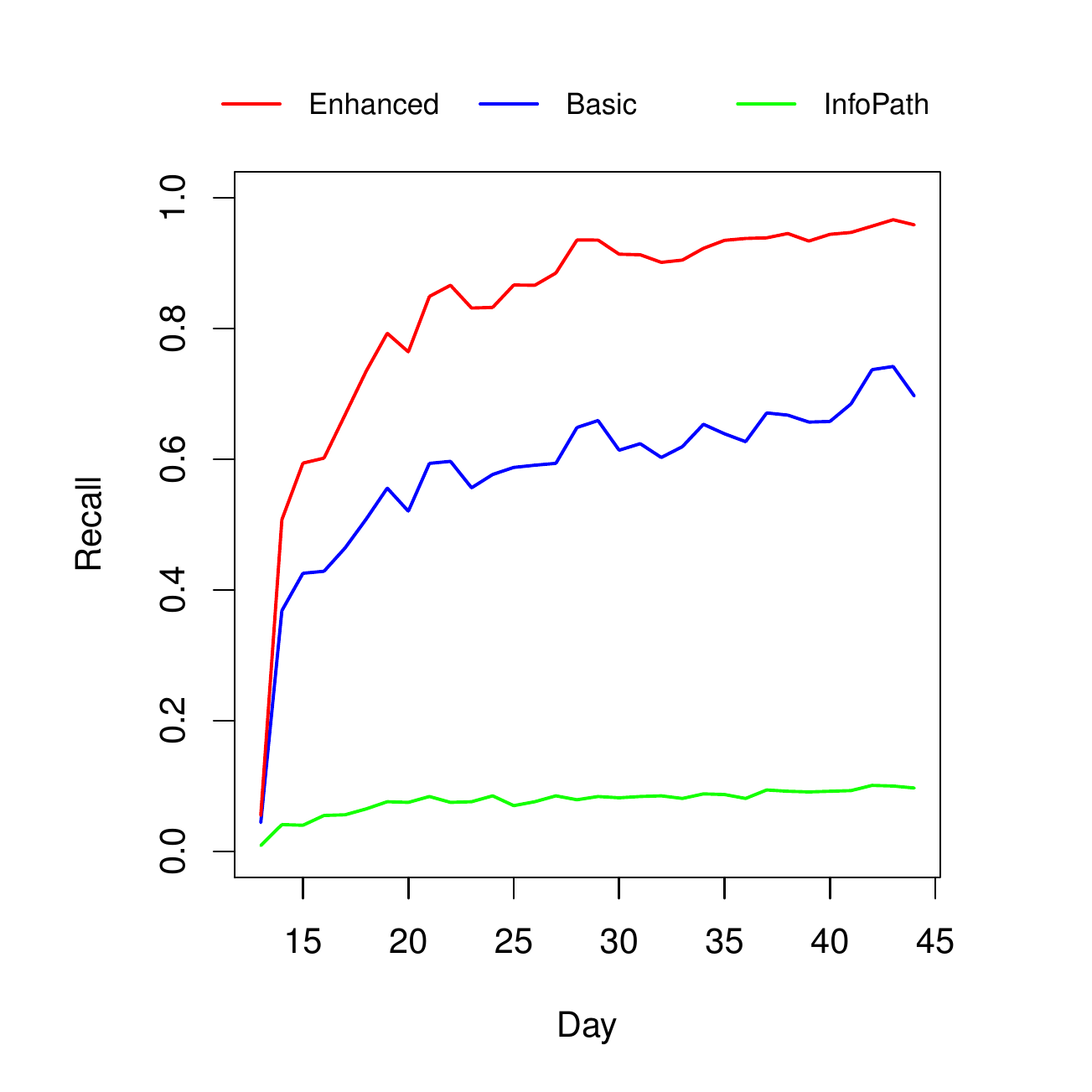}
    \caption{Recall on graph data}
    \label{fig:compare-infopath-dst-recall-g}
  \end{subfigure}
  \begin{subfigure}[t]{0.33\textwidth}
    \includegraphics[width=\textwidth]{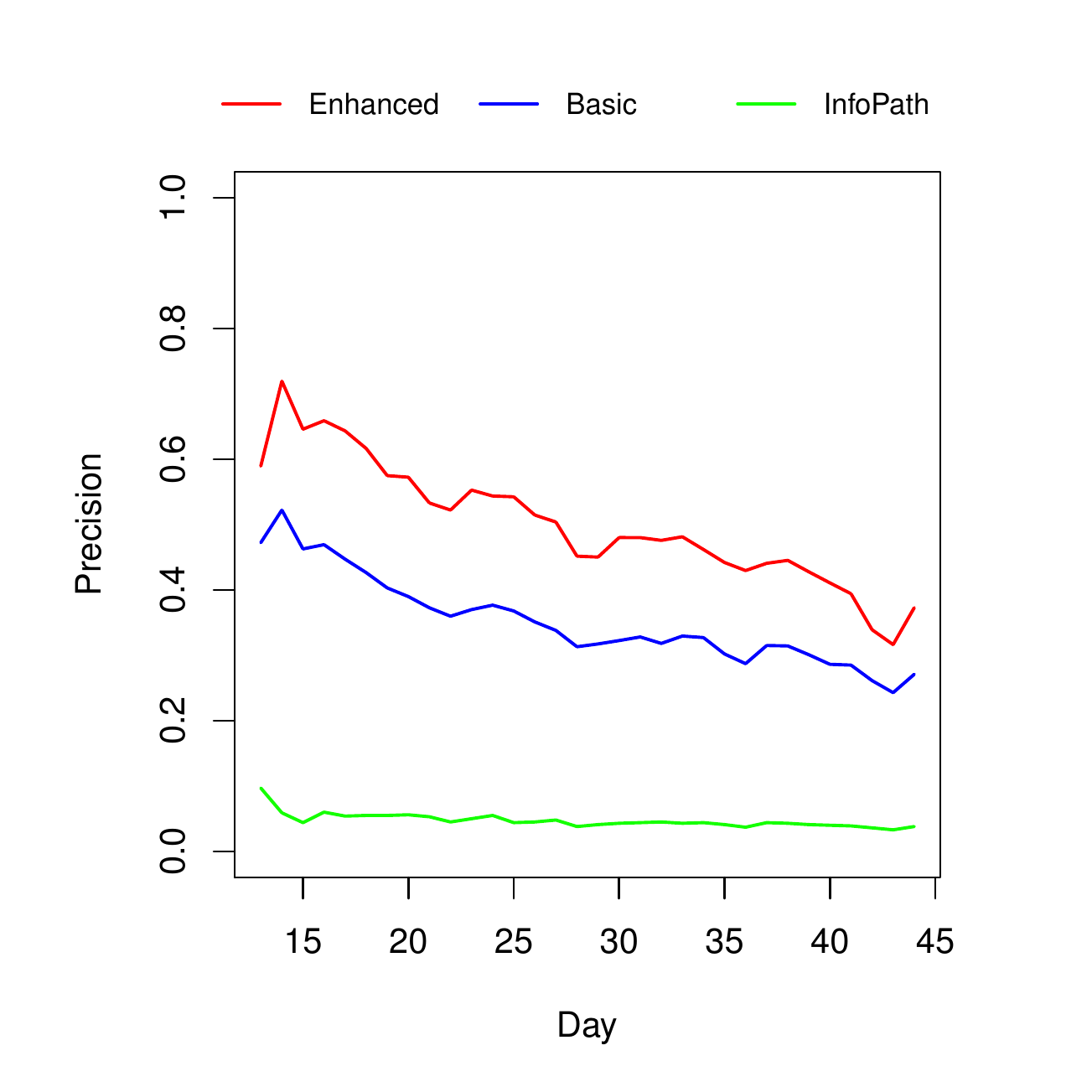}
    \caption{Precision on graph data}
    \label{fig:compare-infopath-dst-prec-g}
  \end{subfigure}
  \begin{subfigure}[t]{0.33\textwidth}
    \includegraphics[width=\textwidth]{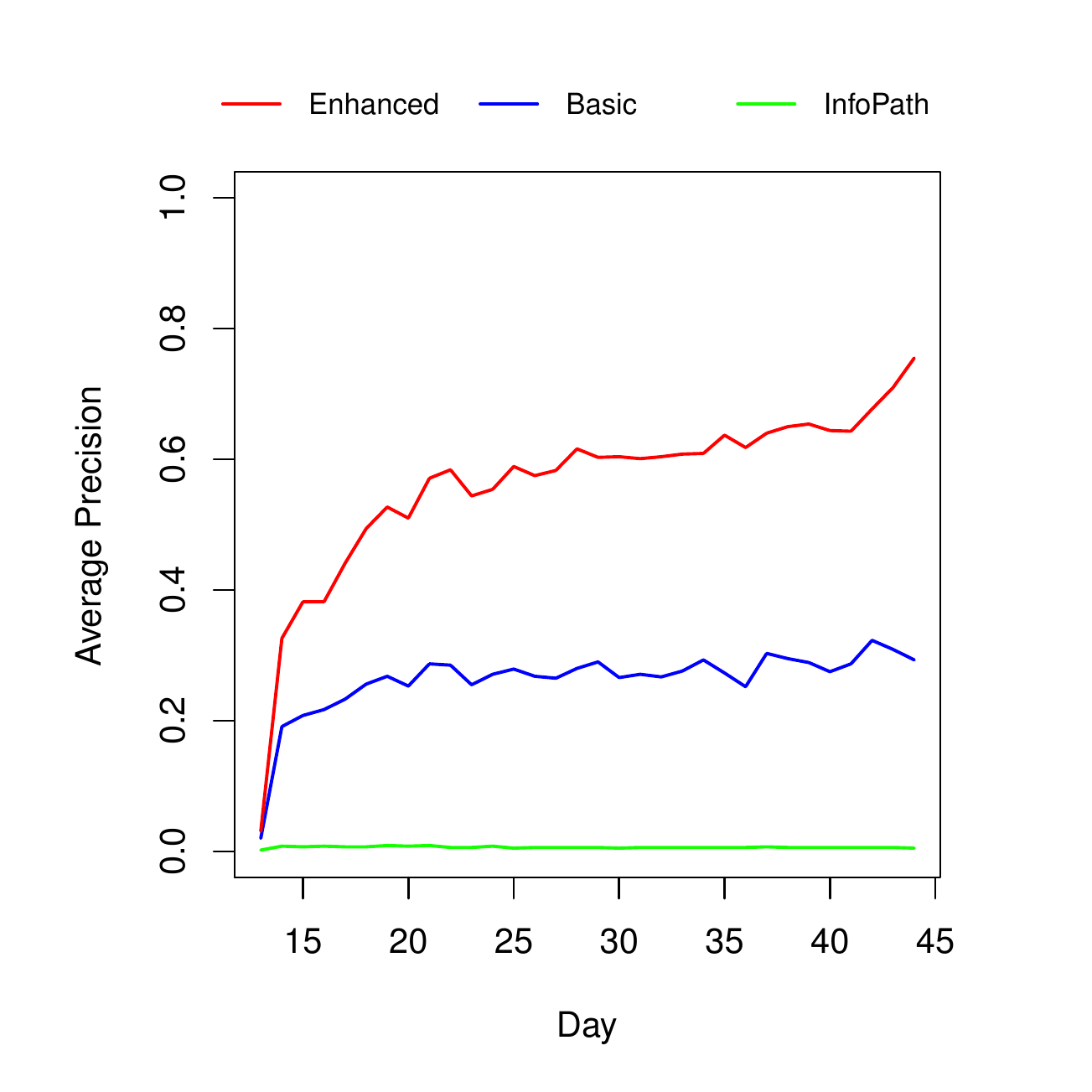}
    \caption{AP on graph data}
    \label{fig:compare-infopath-dst-ap-g}
  \end{subfigure}\\
  \begin{subfigure}[t]{0.33\textwidth}
    \includegraphics[width=\textwidth]{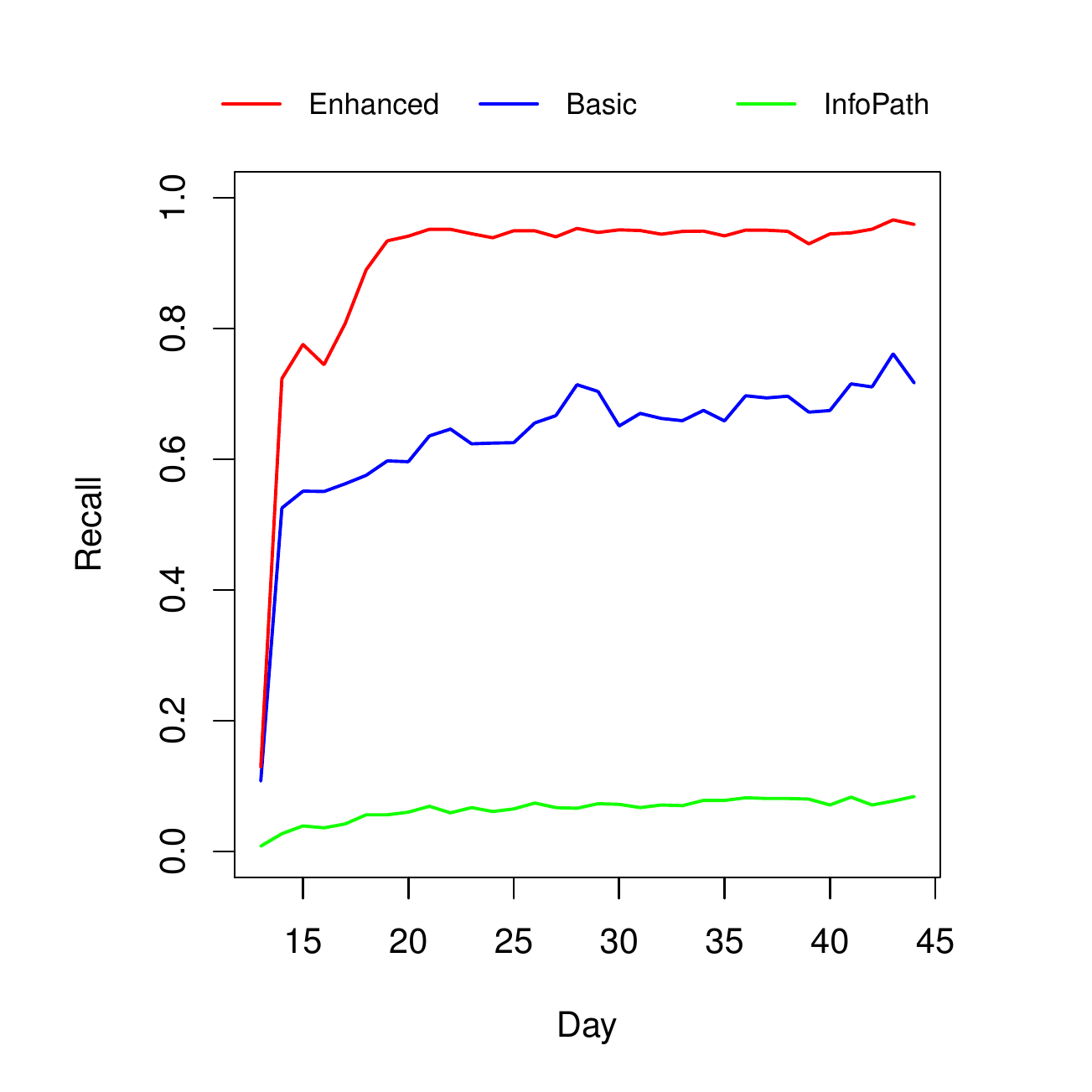}
    \caption{Recall on tree data}
    \label{fig:compare-infopath-dst-recall-t}
  \end{subfigure}
  \begin{subfigure}[t]{0.33\textwidth}
    \includegraphics[width=\textwidth]{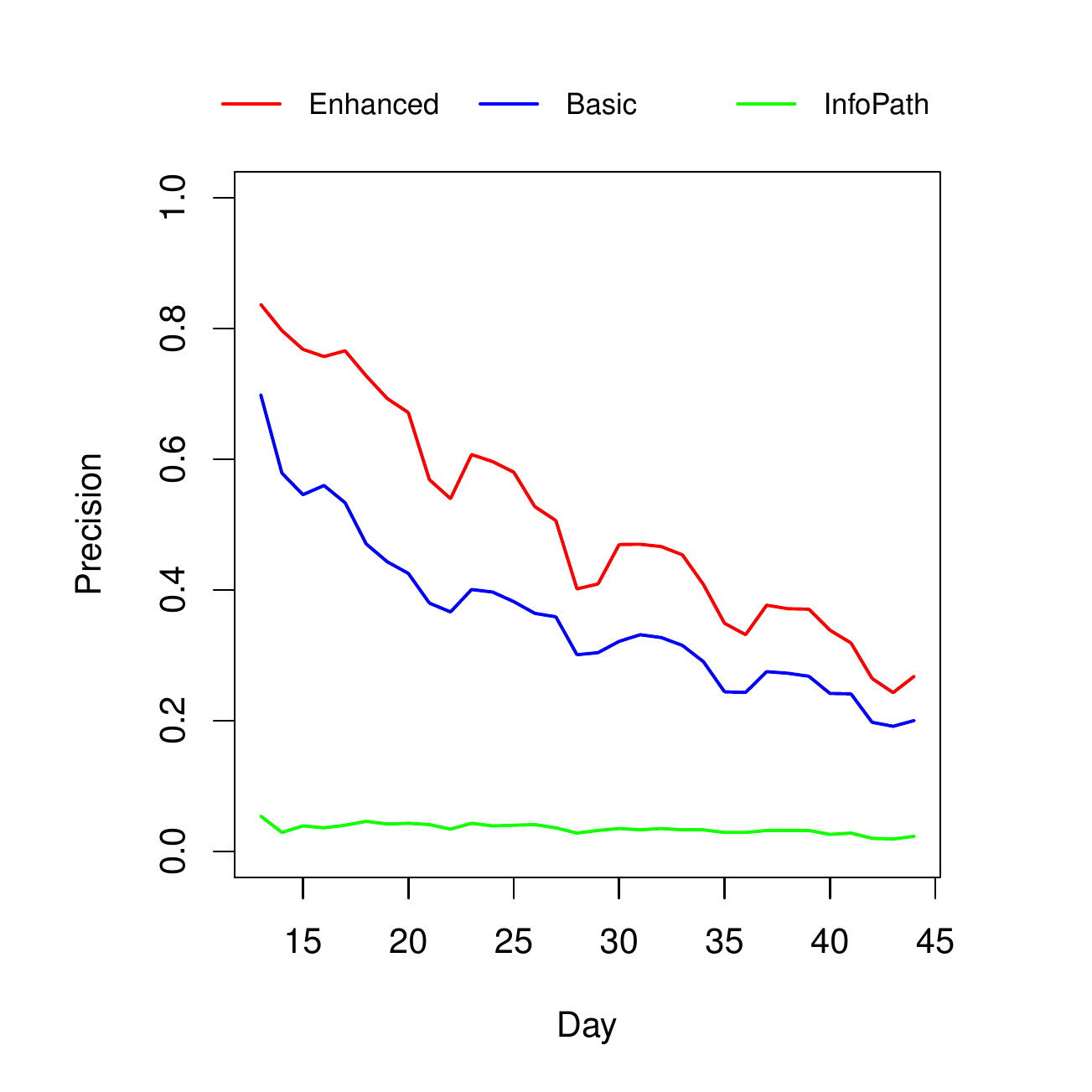}
    \caption{Precision on tree data}
    \label{fig:compare-infopath-dst-prec-t}
  \end{subfigure}
  \begin{subfigure}[t]{0.33\textwidth}
    \includegraphics[width=\textwidth]{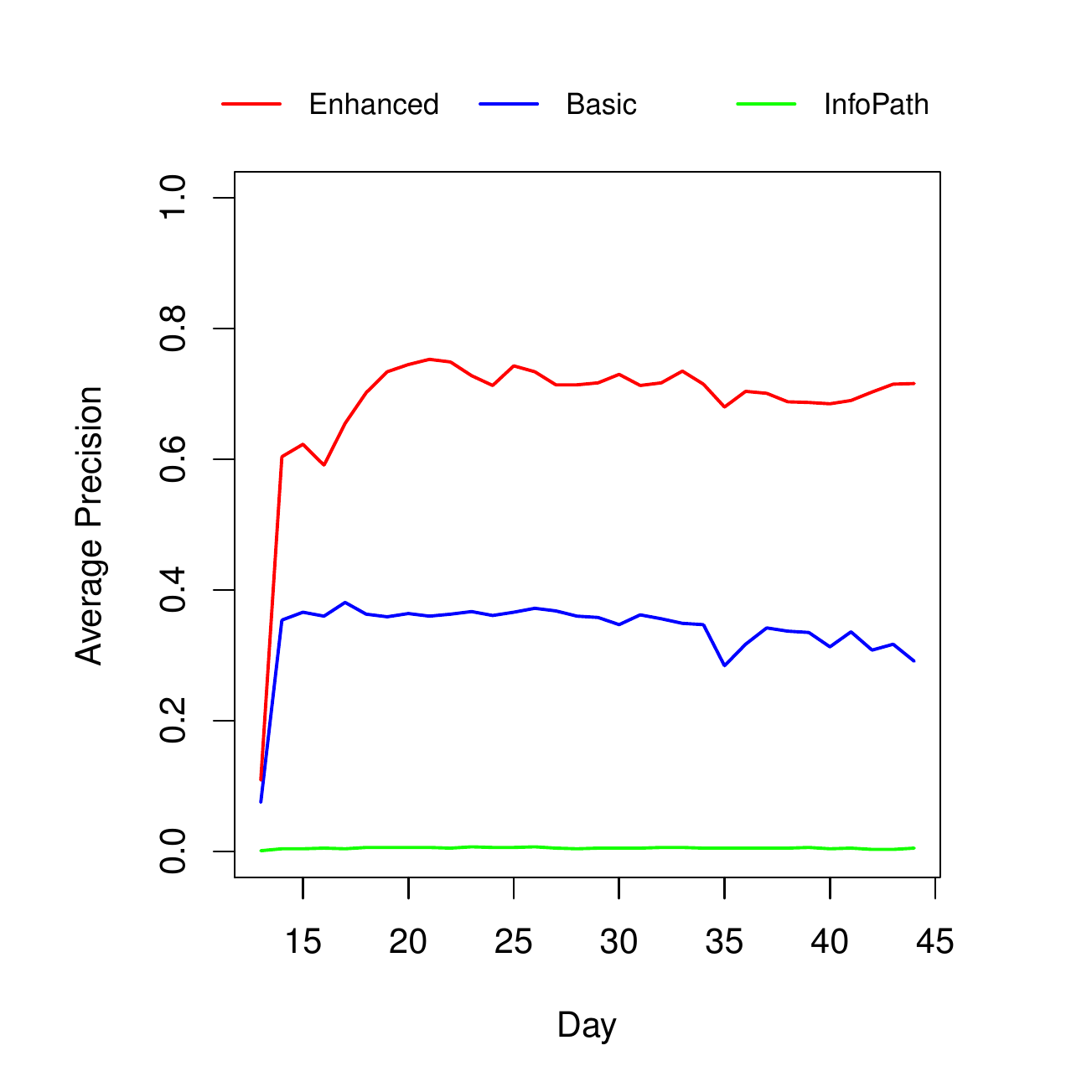}
    \caption{AP on tree data}
    \label{fig:compare-infopath-dst-ap-t}
  \end{subfigure}

  \caption{Recall, precision, and average precision of
    InfoPath and DST on predicting the time-varying networks generated
    per day.  The DST model is trained unsupervisedly on separate cascades
    using basic and enhanced features.  The upper row uses graph-structured
    cascades from the ICWSM 2011 dataset. The lower row uses the subset of
    cascades with tree structures.}
  \label{fig:compare-infopath-dst}
\end{figure*}

\subsection{Feature Sets}
\label{icwsm:feat-sets}

Most existing work on network structure inference described in Related Work
only uses the time difference between two nodes as the feature for learning and
inference. Our model has the ability to incorporate different features as
pointed out in Eq.~\ref{eq:log-linear} and \ref{eq:edge-factored}. Hence in this
paper we experiment different features and report on the following sets:
\begin{itemize}
\item \textit{basic feature sets}, which include only the node
  information and timestamp difference, which resembles what the other
  models do; and
\item \textit{enhanced feature sets}, which include the basic feature
  sets, as well as the languages that both nodes of an edge use, what
  content types as assigned by Spinn3r (blog, news, etc.), whether a
  node is the earliest node in the cluster, and the Jaccard distance
  between the normalized texts in the two nodes.
\end{itemize}
We use one-hot encoding to represent the feature vectors. 
In addition, we discretize real-valued features by binning them.

%
\subsection{Result of Unsupervised Learning at Cascade Level}

In practice, we use Apache Spark for parallelizing the computation to
speed up the optimization process.  We choose batch gradient descent
with a fixed learning rate $5\times 10^{-3}$ and report the result
after $1,500$ iterations.  Inspecting the results of the last two
iterations confirms that all training runs converge. The constraint
set $C$ contains edges that satisfy: (1) time constraints, and (2) only
nodes within the first hour of a specific cascade can be attached to
root.

The DST model outputs finer-grained structure than existing approaches and
predicts a tree for each cascade, with edges equal to the number of nodes.
We report the micro-averaged recall, precision, and F1 for the whole dataset.

The top half of Table~\ref{tbl:cascade-infer-unsup} shows the results
of training the DST model in an unsupervised setting with different
feature sets on both separate cascades dataset and merged cascades
dataset. We also include a naive baseline that simply attaches all
other nodes to the earliest node in a cascade.

From Table~\ref{tbl:cascade-infer-unsup} we can see the flatness
problem.  The naive baseline can already achieve 45\% recall and
58.7\% precision, while knowing the websites and time lags only
yields 34.8\% recall and 45.4\% precision, which partly attributes to
the time constraints we apply on creating the Laplacian matrix so that
the model can at least gets the earliest node and one of the edges
leaving from that node right. On the other hand, the enhanced feature
set utilizes the features from the textual content of posts such as
the Jaccard distance. Having this information helps the DST model
outperform the naive baseline. In the merged clusters setting, instead
of only one seed per cascade being attached to the implicit root, we
have multiple seeds occurring within the same hour attached to the
root. Hence, the naive baseline strategy can at most get the original
cascade to which the earliest node belongs right.  DST with both
feature sets can achieve a better result. We believe in the future,
adding more content based features will further boost the performance.
We expect, however, that disentangling multiple information flow paths
will remain a challenging problem in many domains.

\subsection{Result of Unsupervised Learning at Network Level}

In this section, we evaluate effectiveness on inferring the network
structure, comparing to MultiTree and InfoPath.  The DST model outputs
a tree for each cascade with posterior probabilities for each edge.  To
convert to a network, we sum all posteriors for a certain edge to get
a combined score, from which we obtain a ranked list of edges between
websites.  We report on two different sets of quantitative
measurements: recall/precision/F1 and average precision.

When using InfoPath, we assume an exponential edge transmission model,
and sample 20,000 cascades for the stochastic gradient descent.
The output has the activation rates for a certain edge in the network
per day. We keep those edges which have non-zero activation rate and
actually present on that day to counteract the decaying assumption in
InfoPath. We then compute the recall/precision/average-precision for
each day. To compare the DST model with InfoPath on the time-varying
network, we pick edges from the ranked list of the DST model on each day,
the number of which matches InfoPath's choice. We exclude MultiTree
for the lack of ability to model a dynamic network.

Figure~\ref{fig:compare-infopath-dst} shows the comparison between
InfoPath and the DST model with different feature sets. We can see
that the DST model outperforms InfoPath by a large margin on every
metric with the enhanced feature set being the best.

Now we can compare the DST model with MultiTree and InfoPath on the
static network.  We include every edge in the output of InfoPath.
The top part of Table~\ref{tbl:network-infer} shows a comparison
between the two models in a similar way to the comparisons mentioned
before, where the number of edges from the DST model equals to the
number of total edges selected by InfoPath. As for MultiTree, we keep
all the parameters default while setting the number of edges to match
InfoPath's as well. Since MultiTree assumes a fixed activation rate,
while InfoPath gives activation rate based on the timestep, there
is no way to rank the edges in the static network both methods inferred;
therefore, we don't report average-precision for them.

The DST model also outperforms MultiTree and InfoPath in inferring
static network structure. Notably, the recall/precision of InfoPath is
much higher than the recall/precision per day
(Figure~\ref{fig:compare-infopath-dst}).  This is due to the fact that
edges InfoPath correctly selects in the static network might not be
correct on that specific day.

\subsection{Enforcing Tree Structure on the Data}

In the ICWSM 2011 dataset, 61\% of the cascades are DAGs.  Since DST,
MultiTree, and InfoPath all assume that they are trees, we evaluate
performance on data where this constraint is satisfied---i.e., the
tree-constrained dataset described above.  The bottom part of
Table~\ref{tbl:cascade-infer-unsup} shows that the naive baseline for
separate cascades achieves 94.1\% recall/precision because of
flatness. DST with enhanced features beats it by a mere 0.5\%.  This
leaves very little room for DST to improve in cascade structure
inference problem for separate cascades. For merged cascades, the
naive baseline can at most get the original cascade to which the
earliest node belongs right. DST with basic feature set did adequately
on finding the earliest nodes but found very few correct edges inside
the cascades, while enhanced feature set is better at reconstructing
the cascade structures thanks to the knowledge of textual features,
which leads about a 600\% margin.  With only 24.6\% recall/precision,
there is still room for improvement on this very hard inference
problem.  On network inference, DST with the enhanced feature set also
performs the best for recall and average precision but lags on
precision.  Table~\ref{tbl:network-infer},
Figure~\ref{fig:compare-infopath-dst-recall-t},
\ref{fig:compare-infopath-dst-prec-t} and
\ref{fig:compare-infopath-dst-ap-t} show similar performance when
comparing with MultiTree and InfoPath on inferring different types of
network structure.

\subsection{Result of Supervised Learning at Cascade Level}
\label{sec:supervised}

Our proposed model has the ability to perform both supervised and
unsupervised learning, with different objective functions.  One of the
main contributions of the DST model is to be able to learn the cascade
level structure in a feature-enhanced and unsupervised way. However,
supervised learning can establish on upper bound for unsupervised
performance when trained with the same features.

Table~\ref{tbl:sup-merged} shows the result of supervised learning
using DST on the merged cascades with tree structure enforced. Since
there are only 938 merged cascades, we perform a 10-fold cross
validation on both dataset and we report the result of 5 folds. We
split the training and test set by interleaved round-robin sampling
from the merged cascades dataset.  Although not precisely comparable
to DST in the unsupervised setting due to this jackknifing,
Table~\ref{tbl:sup-merged} still shows results about twice as large as
for unsupervised training.

\section{Conclusion}
\label{conclusion}

We have proposed a method to uncover the network structure of
information cascades using an edge-factored, conditional log-linear
model, which can incorporate more features than most comparable
models. This directed spanning tree (DST) model can also infer finer
grained structure on the cascade level, besides inferring global
network structure.  We utilize the matrix-tree theorem to prove that
the likelihood function of the conditional model can be solved in
cubic time and to derive a contrastive, unsupervised training
procedure.  We show that for ICWSM 2011 Spinn3r dataset, our proposed
method outperforms the baseline MultiTree and InfoPath methods in
terms of recall, precision, and average precision.  In the future, we
expect that applications of this technique could benefit from richer
textual features---including full generative models of child document
text---and different model structures trained with the contrastive
approach.

\section{Acknowledgements}

This research was supported by the National Institute on Deafness and
Other Communication Disorders of the National Institutes of Health
under award number R01DC009834 and by the Andrew W. Mellon Foundation.
The content is solely the responsibility of the authors and does not
necessarily represent the official views of the NIH or the Mellon.

\bibliography{Xu-Smith}
\bibliographystyle{aaai}
\end{document}